\documentclass[pra,aps,amsfonts,amsmath,twocolumn,showpacs,floatfix,superscriptaddress]{revtex4}
\usepackage{bm}
\usepackage{amsfonts}
\usepackage{amsmath}
\usepackage{graphicx}
\newcommand{\vek}[1]{\bm{\mathrm{#1}}}

\newcommand{\calF}{\mathcal{F}}
\newcommand{\calG}{\mathcal{G}}
\newcommand{\jv}{\vek{j}}
\newcommand{\Lv}{\vek{L}}
\newcommand{\loc}{\mathit{loc}}
\newcommand{\nablav}{\vek{\nabla}}
\newcommand{\Omegav}{\vek{\Omega}}

\newcommand{\pv}{\vek{p}}
\newcommand{\rv}{\vek{r}}
\newcommand{\vv}{\vek{v}}
\newcommand{\Eq}[1]{Eq.\@ (\ref{#1})}
\newcommand{\Eqs}[1]{Eqs.\@ (\ref{#1})}
\newcommand{\Ref}[1]{Ref.\@ \cite{#1}}
\newcommand{\Fig}[1]{Fig.\@ \ref{#1}}

\begin{document}
\title{Pair Breaking in Rotating Fermi Gases}
\author{Michael Urban}
\affiliation{Institut de Physique Nucl{\'e}aire, CNRS-IN2P3 and
Universit\'e Paris-Sud, 91406 Orsay Cedex, France}
\author{Peter Schuck}
\affiliation{Institut de Physique Nucl{\'e}aire, CNRS-IN2P3 and
Universit\'e Paris-Sud, 91406 Orsay Cedex, France}
\affiliation{Laboratoire de Physique et Mod\'elisation des Milieux
Condens\'es, CNRS and Universit\'e Joseph Fourier, Maison des
Magist\`eres, BP 166, 38042 Grenoble Cedex, France}
\begin{abstract}
We study the pair-breaking effect of rotation on a cold Fermi gas in
the BCS-BEC crossover region. In the framework of BCS theory, which is
supposed to be qualitatively correct at zero temperature, we find that
in a trap rotating around a symmetry axis, three regions have to be
distinguished: (A) a region near the rotational axis where the
superfluid stays at rest and where no pairs are broken, (B) a region
where the pairs are progressively broken with increasing distance from
the rotational axis, resulting in an increasing rotational current,
and (C) a normal-fluid region where all pairs are broken and which
rotates like a rigid body. Due to region B, density and current do not
exhibit any discontinuities.
\end{abstract}
\maketitle
The surprising properties of superfluids become most evident if one
looks at rotating systems. But the rotation does not only reveal the
superfluidity, it can also destroy it. To give an example, in nuclear
physics, the strong reduction of the nuclear moment of inertia
compared to its rigid-body value is a direct consequence of
superfluidity due to pairing correlations. But with increasing angular
momentum, the pairing correlations are progressively destroyed and the
moment of inertia increases to its rigid-body value. This
pair-breaking effect of rotation was studied many years ago
\cite{nuclei}.

In trapped atomic Fermi gases, the picture is somewhat different,
since, contrary to the situation in atomic nuclei, the coherence
length is much smaller than the system size. It is therefore possible
to create quantized vortices or even vortex lattices \cite{Zwierlein},
which allow the system to stay superfluid while rotating.

However, in a recent paper by Bausmerth, Recati, and Stringari
\cite{Bausmerth} it has been argued that it may be possible to put a
trapped Fermi gas adiabatically into rotation without creating
vortices. In that paper, the destruction of superfluidity by rotation
is described in a way which is very different from the nuclear physics
case: Instead of decreasing the value of the pairing gap with
increasing angular velocity, the authors assume that the system
separates into a paired and an unpaired phase, while the properties of
the paired phase itself are not affected by the rotation. The authors
consider the unitary limit, where the energy densities of the paired
and unpaired phases are known from Quantum-Monte-Carlo (QMC)
simulations \cite{QMC}. The phase boundary between the paired and the
unpaired phases is determined by energy minimization: Near the
rotational axis, the system prefers to stay superfluid, i.e., to stay
at rest, since the paired phase has a lower energy density than the
unpaired one. But beyond a certain distance from the rotational axis,
the centrifugal energy which the system could win if it participated
in the rotation becomes equal to the energy which is needed to break
the pairs. Hence, the non-rotating superfluid core is surrounded by a
rotating normal-fluid phase. At the interface separating the two
phases, the density and the current are discontinuous.

This picture is very intuitive, but it is lacking the microscopic
understanding of the pair-breaking mechanism. In the present paper we
will therefore describe the rotating Fermi gas in the framework of BCS
theory. The rotation is most easily described in the rotating frame,
where the hamiltonian $\hat{H}$ (minus the chemical potential $\mu$
times the particle number $\hat{N}$) is given by
\begin{multline}
\hat{H}-\mu \hat{N} = \int d^3r \Big[\hat{\psi}^\dagger(\rv)
  \Big(\frac{\pv^2}{2m}+V(\rv)-\Omega L_z-\mu\Big)\hat{\psi}(\rv)\\
  +g \hat{\psi}^\dagger_{\uparrow}(\rv) \hat{\psi}^\dagger_{\downarrow}(\rv)
  \hat{\psi}_{\downarrow}(\rv)\hat{\psi}_{\uparrow}(\rv)\Big]\,,
\end{multline}
where $\hat{\psi}$ is the Fermion field operator with components for
(pseudo-)spin up ($\uparrow$) and down ($\downarrow$), $m$ is the atom
mass, $\pv = -i\hbar\nablav$ and $\Lv = \rv\times\pv$ are momentum and
angular momentum, respectively, $V(\rv) = m (\omega_z^2
z^2+\omega_\perp^2 r_\perp^2)/2$ is the axially symmetric trap
potential and $g < 0$ is the coupling constant. The system is supposed
to rotate with angular velocity $\Omega$ around the symmetry ($z$)
axis of the potential.

If the system is large enough, such that the coherence length is small
compared with the oscillator length associated with the trap
potential, we can make use of the Thomas-Fermi (TF) or local-density
approximation (LDA), which amounts to treating the system at each
point $\rv$ as uniform with a local chemical potential $\mu_\loc(\rv)
= \mu - V(\rv)$. Then $\pv$ becomes a number instead of an operator,
and the ``cranking'' term $\Omega L_z$ can conveniently be written as
$\Omega L_z = \vv(\rv)\cdot \pv$, where $\vv(\rv) = \Omegav\times\rv$
is the velocity field corresponding to a rigid rotation. All
quantities depend only parametrically on $\rv$ via $\mu_\loc(\rv)$ and
$\vv(\rv)$.

The gap, density, and current can all be derived from the normal and
anomalous Matsubara Green's functions $\calG$ and $\calF^\dagger$
\cite{FetterWalecka}. They have to satisfy the Gorkov equations, which
in the presence of the cranking term $\Omega L_z$ become
\begin{align}
(i\hbar\omega_n - \xi + \vv\cdot\pv) \calG 
  + \Delta \calF^\dagger &= \hbar\,,
\label{Gorkovequation1}
\\
(i\hbar\omega_n + \xi + \vv\cdot\pv) \calF^\dagger
  + \Delta^* \calG &= 0\,.
\label{Gorkovequation2}
\end{align}
where we introduced the abbreviation $\xi = \xi(\rv,\pv) =
p^2/(2m)-\mu_\loc(\rv)$, $\omega_n$ denotes a fermionic Matsubara
frequency, and $\Delta(\rv)$ is the gap. Note that we are neglecting
the Hartree mean field, but anyway it would not qualitatively change
our results in the BCS-BEC crossover regime
\cite{Perali}. \Eqs{Gorkovequation1} and (\ref{Gorkovequation2}) can
readily be solved for $\calG$ and $\calF$. They are formally similar
to those describing pairing between particles with unbalanced
populations (see, e.g., \Ref{Lombardo}), except that here the chemical
potentials for the two spins are equal and the asymmetry is between
states with opposite momenta ($\pv$ and $-\pv$).

In the case of a system without superfluid flow (like in our axially
symmetric trap, as long as there are no vortices), the gap can be
assumed to be real ($\Delta = \Delta^*$). The gap equation is obtained
in the usual way by summing $\calF$ over $\omega_n$ and integrating
over $\pv$, with the result
\begin{equation}
\Delta = -\frac{4\pi\hbar^2 a}{m}\int\frac{d^3p}{(2\pi\hbar)^3}
\Big(\frac{\Delta}{2E}[1-f(E_+)-f(E_-)]-\frac{m\Delta}{p^2}\Big)\,,
\label{gapequation2}
\end{equation}
where we defined the quasiparticle energies $E_\pm = E\pm\pv\cdot\vv$,
with $E = \sqrt{\xi^2+\Delta^2}$, and $f(E) = 1/(e^{E/(k_B T)}+1)$
denotes the Fermi function, $T$ being the temperature and $k_B$ the
Boltzmann constant. In \Eq{gapequation2}, the divergence of the gap
equation due to the contact interaction has been regularized in the
usual way by expressing the coupling constant $g$ in terms of the
$s$-wave scattering length $a$ \cite{SadeMelo}.

We are mainly interested in the BCS-BEC crossover regime, where it is
known that the BCS description fails at higher temperatures, and in
particular the BCS prediction for the critical temperature $T_c$ is
much too high. However, at zero temperature, BCS theory gives a
reasonable description throughout the crossover. We will therefore
restrict ourselves to the zero-temperature case, in which the Fermi
function reduces to a step function, $f(E) = \theta(-E)$. Hence, the
factor $[1-f(E_+)-f(E_-)]$ is equal to $1$ if both $E_+$ and $E_-$ are
positive and $0$ otherwise (at most one of the two energies $E_+$ and
$E_-$ can be negative). In other words, states with $E_\pm < 0$ are
excluded from pairing. In order to better understand the role of these
states, let us look at the occupation numbers $\rho(\rv,\pv)$, which
are obtained by summing $\calG$ over $\omega_n$:
\begin{equation}
\rho(\rv,\pv) = \frac{1}{2}\Big(1-\frac{\xi}{E}\Big)[1-f(E_+)]
   +\frac{1}{2}\Big(1+\frac{\xi}{E}\Big)f(E_-)\,.
\label{occupation}
\end{equation}
For states with both $E_+ > 0$ and $E_- > 0$, this reduces to the
usual BCS expression. But if a state with momentum $\pv$ has $E_- <
0$, its occupation number is equal to $1$. The corresponding
time-reversed state with momentum $-\pv$ has then $E_+ < 0$ and its
occupation number is equal to $0$. As we will see below, this gives
rise to a normal-fluid (rotational) current.

It is easy to see that the energies $E_\pm$ can only become negative
if the velocity $v$ exceeds a critical value such that
\begin{equation}
p_F^\prime v > \Delta\,.
\label{condition1}
\end{equation}
Here we have introduced the abbreviation $p_F^\prime =
\sqrt{2m\mu_\loc^\prime}$, where $\mu_\loc^\prime = \mu_\loc+mv^2/2$
denotes the local chemical potential which includes the effect of the
centrifugal force, and $p_F^\prime$ is the corresponding local Fermi
momentum. For a given $z$ coordinate, the condition (\ref{condition1})
is fulfilled beyond a certain distance $r_{\perp 1}(z)$ from the
rotational axis, since the velocity increases as $v = \Omega
r_\perp$. At smaller distances, the energies $E_\pm$ are always
positive, i.e., the system is in the usual superfluid phase and does
not participate in the rotation. Beyond $r_{\perp 1}$, the gap is
reduced by the rotation. We will call this region, where a rotational
current exists although the gap is non-zero, the partially paired
phase. Finally, at a certain distance $r_{\perp 2}$, the gap vanishes
and the system enters the normal phase where it rotates like a rigid
body.

If the condition (\ref{condition1}) is fulfilled, i.e., for $r_\perp >
r_{\perp 1}$, one can easily see that the energies $E_\pm$ can become
negative if the momentum lies between two limits $p_-$ and $p_+$ which
are given by
\begin{equation}
p_\pm^2 = p_F^{\prime\,2}+m^2 v^2\pm 2m\sqrt{p_F^{\prime\,2}
  v^2-\Delta^2}\,,
\end{equation}
The integrand of the gap equation (\ref{gapequation2}) is only
affected by the rotation if $p$ lies between $p_-$ and
$p_+$. Integrating \Eq{gapequation2} over the angle between $\pv$ and
$\vv$ and dividing both sides of the equation by $\Delta$, we obtain
\begin{equation}
1 = -\frac{a}{\pi\hbar m}\Big[
  \int_0^\infty dp \Big(\frac{p^2}{E}-2m\Big)-
  \int_{p_-}^{p_+} dp \Big(\frac{p^2}{E}-\frac{p}{v}\Big)\Big]\,.
\label{gapequation3}
\end{equation}
The first integral is the same as in the gap equation without rotation
while the second one is the contribution of the $f(E_\pm)$ terms due
to the rotation.

In the weak-coupling limit, when $\Delta\ll\mu_\loc$, the
pair-breaking effects appear already at extremely low angular
velocities $\Omega$. In this case it is possible to evaluate the
integrals in \Eq{gapequation3} analytically, and one can show that the
critical velocity for which the gap disappears is given by $v_c =
(e/2)\Delta_{v=0}/p_F$, where $e = 2.71\dots$ denotes Euler's
number. Hence, for a given $z$ coordinate, the radial coordinates
$r_{\perp 1,2}$ separating the fully paired from the partially paired
and the partially paired from the unpaired phase, respectively, are
the solutions of the equations
\begin{align}
p_F(r_{\perp 1},z) \Omega r_{\perp 1} &=
  \Delta_{\Omega=0}(r_{\perp 1},z)\,,
  \label{rperp1}\\ 
p_F(r_{\perp 2},z) \Omega r_{\perp 2} &=
  \frac{e}{2}\Delta_{\Omega=0}(r_{\perp 2},z)\,.
  \label{rperp2}
\end{align}

In the crossover regime, the situation is more complicated, since the
gap $\Delta$ may be comparable with $\mu_\loc$. Therefore the
integrals have to be evaluated numerically. In addition, the rotation
can now be much faster and the centrifugal force can lead to a
sizeable change of the density profile and it is necessary to readjust
the global chemical potential $\mu$ as a function of $\Omega$ in order
to keep the total number of particles fixed. The density per spin
state, $\rho(\rv)$, is obtained by integrating the occupation numbers
over $\pv$. Using \Eq{occupation}, one obtains
\begin{equation}
\rho(\rv) = \frac{1}{4\pi^2\hbar^3}\Big[
    \int_0^\infty dp\,p^2\Big(1-\frac{\xi}{E}\Big)
   +\int_{p_-}^{p_+} dp\,\xi\Big(\frac{p^2}{E}-\frac{p}{v}\Big)\Big]\,.
\label{density1}
\end{equation}
The second term arises from the $f(E_\pm)$ terms and exists only if
the condition (\ref{condition1}) is fulfilled, i.e., beyond $r_{\perp
1}$. Between $r_{\perp 1}$ and $r_{\perp 2}$, the density goes
smoothly from its value with pairing to the value without pairing,
$\lim_{\Delta\to 0}\rho(\rv) =
p_F^{\prime\,3}(\rv)/(6\pi^2\hbar^3)$. Once we have calculated the
density, we can obtain the total number of particles by integrating
the density over space. This allows us to determine the value of the
chemical potential.

An interesting quantity is the current density, which can be obtained
by multiplying the occupation numbers with $\pv/m$ and integrating
over $\pv$. From \Eq{occupation} it is clear that for
$r_\perp<r_{\perp\,1}$, i.e., close to the rotational axis where the
condition (\ref{condition1}) is not satisfied, the current vanishes as
it should in the superfluid phase.  Beyond $r_{\perp 1}$, the result
can be given in closed form as
\begin{equation}
\jv =\frac{(p_F^{\prime\,2}-\Delta^2/v^2)^{3/2}}{6\pi^2\hbar^3}\vv\,.
\label{current}
\end{equation}
One sees that in the partially paired phase the current increases with
decreasing gap and it correctly approaches its rigid-body limit if one
approaches the unpaired phase: $\lim_{\Delta\to 0} \jv(\rv) =
\rho(\rv) \vv(\rv)$.

\begin{figure}
\includegraphics[width=7.3cm]{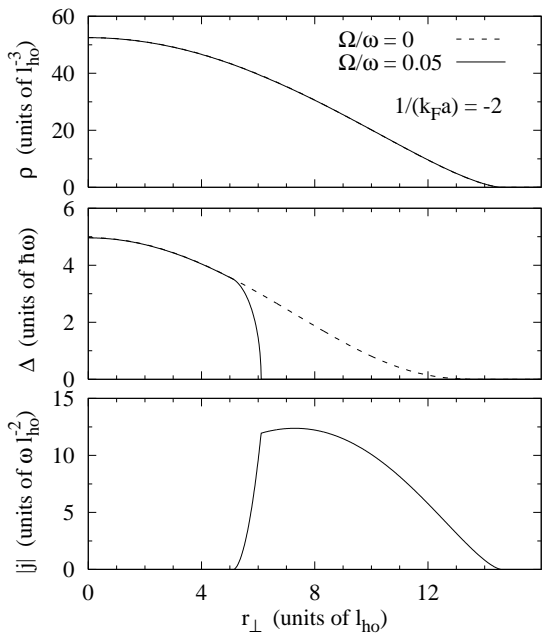}
\caption{From top to bottom: density per spin state $\rho$, gap
$\Delta$, and current $|\jv|$ in a rotating Fermi gas ($4\cdot 10^5$
atoms in an isotropic trap with frequency $\omega$) in the BCS phase
as a function of the distance $r_\perp$ from the $z$ axis, for $z =
0$. The solid lines correspond to a gas rotating with angular velocity
$\Omega = 0.05\omega$. For comparison, the results for the
non-rotating case (ground state) are shown as the dashed lines.}
\label{figbcs}
\end{figure}
\begin{figure}
\includegraphics[width=7.3cm]{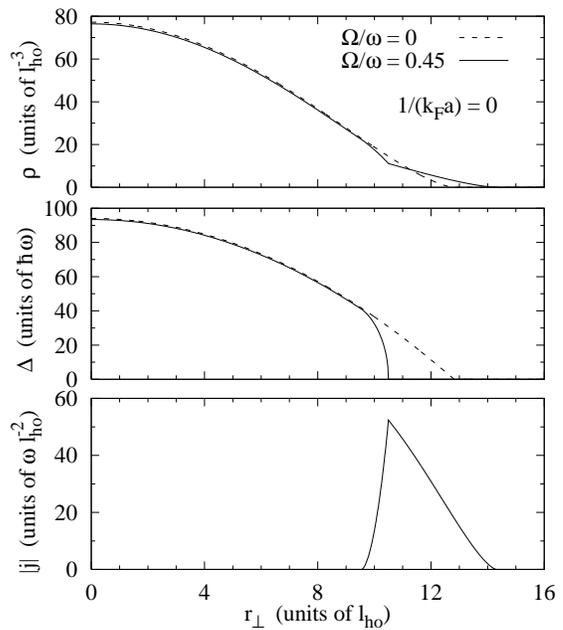}
\caption{Same as \Fig{figbcs}, but for a unitary Fermi gas rotating
with angular velocity $\Omega = 0.45\omega$.}
\label{figunitary}
\end{figure}
Let us now discuss some numerical results. We consider a system with
$N = 4\cdot 10^5$ atoms ($2\cdot 10^5$ atoms per spin state) in two
cases: (a) close to the BCS limit, with $1/(k_F a) = -2$ [$k_F =
p_F(\rv = 0)/\hbar$], and (b) at unitarity, i.e., in the limit $a\to
\infty$.  We do not consider the BEC side of the cross-over, since as
soon as the chemical potential becomes negative, the energies $E_\pm$
are always positive, i.e., the molecules in the BEC phase are never
broken by the rotation. For simplicity we choose a spherically
symmetric trap ($\omega_z = \omega_\perp$), but this will not
qualitatively change our results. In the figures, we will use the
harmonic oscillator units set by the trap potential, i.e.,
$\hbar\omega$ for energies and $l_{ho} = \sqrt{\hbar/(m\omega)}$ for
lengths.

Let us first discuss the BCS case. In this case the pairing is so weak
that it does not appreciably influence the density (upper panel of
\Fig{figbcs}). It is also very fragile, i.e., the moment of inertia,
which can be calculated within linear response theory
\cite{UrbanSchuck}, must be measured at extremely low angular
velocity. Already for an angular velocity as small as $\Omega =
0.05\omega$, the gap (second panel of \Fig{figbcs}) is zero in a large
part of the system. Because of the small angular velocity, the
centrifugal force has no effect on the density, either. Looking at the
gap, one can clearly see the point $r_{\perp 1}(z=0) = 5.1\,l_{ho}$
where the results for the non-rotating (dashed line) and the rotating
(solid line) system start to differ, and the point $r_{\perp 2} =
6.1\, l_{ho}$ where the gap goes to zero. The three regions are even
more evident in the current (lower panel of \Fig{figbcs}): The current
starts to be non-vanishing at $r_{\perp 1}$ and it has a kink at
$r_{\perp 2}$ where it reaches the rigid-body value.

More interesting are the results in the cross-over regime, where the
gap is strong enough to support a relatively fast rotation. In
\Fig{figunitary} we display the density, gap and current (from top to
bottom) for a system at the unitary limit rotating with $\Omega =
0.45\,\Omega$ (solid lines; for comparison, the density and gap of the
corresponding non-rotating system are shown as the dashed lines). In
this case, the centrifugal force leads to an oblate deformation of the
system: The chemical potential $\mu$ and the axial size of the system,
which is determined by $z_\mathit{max} = \sqrt{2\mu/m}/\omega$,
decrease (in the present example, $\mu$ decreases from 81.7 to 81.2
$\hbar\omega$), while the radial size, which is determined by
$r_{\perp \mathit{max}} = \sqrt{2\mu/[m(\omega^2-\Omega^2)]}$,
increases. The increase of the radial size is visible in the upper
panel of \Fig{figunitary}, where the density is shown as a function of
$r_\perp$ for $z = 0$. The depletion of the density in the center is a
consequence of the reduced chemical potential. This is also the reason
why the gap in the center decreases with the rotation. Due to the
strong pairing, the gap has a direct effect on the density. This is
the reason for the kink in the density profile at $r_\perp = r_{\perp
2}$. However, we stress that the density stays continuous at $r_{\perp
2}$.

The fact that, in contrast to the results of Bausmerth et
al. \cite{Bausmerth}, the density, the gap, and the current remain
continuous functions of $r_\perp$ is the main statement of the present
paper. In fact, if we followed the arguments given in \Ref{Bausmerth},
we would find a similar discontinuity as they do. The only difference
with their result would be the different numerical values of the
parameters $\xi_S$ and $\xi_N$ which determine the relationship
between the density and the local chemical potential [$\mu_\loc =
\xi_S \hbar^2 (6\pi^2\rho)^{2/3}/(2m)$ or $\mu_\loc^\prime =
\xi_N \hbar^2 (6\pi^2\rho)^{2/3}/(2m)$ in the superfluid and
normal phase, respectively]. In BCS theory without mean-field shift,
one obtains $\xi_S = 0.59$ and $\xi_N = 1$, whereas the QMC results
used in \Ref{Bausmerth} are $\xi_S = 0.44$ and $\xi_N = 056$. If one
excluded the possibility of an intermediate ``partially paired''
phase, as in \Ref{Bausmerth}, the system would have to split into a
fully paired superfluid and a fully unpaired normal-fluid phase, and
the density would have a discontinuity across the phase boundary with
a ratio $\rho_N / \rho_S = (\xi_S/\xi_N)^{3/5}$, which gives $0.73$
with the BCS results and $0.85$ with the QMC results for $\xi_S$ and
$\xi_N$. From this we see that, even if BCS theory is not capable to
give the right numbers for $\xi_S$ and $\xi_N$, the ratio is
semi-quantitatively correct. Anyway, even if our results for the
unitary limit might not be very precise, we believe that they are
qualitatively correct and that between the ordinary normal and
superfluid phases there will be a region in which some pairs are
broken while others stay unbroken. In particular, we checked that in
the region between $r_{\perp 1}$ and $r_{\perp 2}$ our energy density
is lower than both that of the non-rotating superfluid and the ridigly
rotating unpaired gas.

We emphasize that the existence of the intermediate region is not a
finite-size effect, but it survives in arbitrarily large systems. For
instance, if the trap was a flat potential well instead of a harmonic
oscillator, the ratio of the two radii $r_{\perp 2}$ and $r_{\perp 1}$
would become $r_{\perp 2}/r_{\perp 1} = e/2 = 1.36$ according to
\Eqs{rperp1} and (\ref{rperp2}), independently of the size of the
system and of the angular velocity of the rotation. In the harmonic
oscillator the intermediate region is smaller since the gap decreases
with increasing $r_\perp$ already in the non-rotating case.

This does not mean that finite-size effects do not play any role. For
instance, the abrupt decrease of $\Delta$ for $r_\perp \to r_{\perp
2}$ is an artefact of the TF approximation, which requires that all
spatial variations be slow compared with the length scale set by the
coherence length. A necessary condition for this is $\Delta \gg \hbar
\omega$. In a true quantum calculation, the profiles of $\Delta$,
$\rho$ and $|\jv|$ would be rounded and no sharp interface between the
different phases could be defined. In addition, beyond a certain
critical angular velocity $\Omega_c$ the gap should completely
disappear, even on the rotational axis \cite{ZhaiHo}.

An interesting extension of the present work is to study a system
which is deformed in the $xy$ plane, i.e., in the plane perpendicular
to the rotational axis. This question is very important since it is
impossible to put the system into rotation without such a deformation
(of course, once the system rotates, the deformation can be switched
off and the conservation of angular momentum ensures that the system
keeps rotating). In the deformed case, also the superfluid part of the
system has a non-vanishing current, with an irrotational velocity
field. Another important question concerns the collective excitations
of the rotating system, in particular the radial quadrupole mode whose
precession is used in current experiments for measuring the angular
momentum of the system \cite{Riedl}. In order to stay in contact with
the experiments, temperature effects should be taken into account,
too.

\end{document}